\documentclass[12pt,letter]{article}
\usepackage{graphics}
\usepackage{graphicx}
\usepackage[latin1]{inputenc}
\usepackage{amsmath}
\usepackage{amssymb}

\newcommand{\Lyx}{L\kern-.1667em\lower.25em\hbox{y}\kern-.125emX\spacefactor1000}

\newcommand{\erfc}{\mathop{\text{erfc}}}

\begin{document}
\bibliographystyle{plain} 
\pagestyle{plain} 
\pagenumbering{arabic}
\title{Wigner Surmise For Domain Systems}
\author {Diego Luis Gonz\'alez\footnote{die-gon1@uniandes.edu.co}\\
        Gabriel T\'ellez\footnote{gtellez@uniandes.edu.co}\\
        Departamento de F\'{\i}sica, Universidad de Los Andes\\
        A.~A.~4976 Bogot\'a, Colombia.}

\date{}
\maketitle

\begin{abstract}
In random matrix theory, the spacing distribution functions $p^{(n)}(s)$ are well fitted by the Wigner surmise and its generalizations. In this approximation the spacing functions are completely described by the behavior of the exact functions in the limits $s\rightarrow0$ and $s\rightarrow\infty$. Most non equilibrium systems do not have analytical solutions for the spacing distribution and correlation functions. Because of that, we explore the possibility to use the Wigner surmise approximation in these systems. We found that this approximation provides a first approach to the statistical behavior of complex systems, in particular we use it to find an analytical approximation to the nearest neighbor distribution of the annihilation random walk.
\end{abstract} 

{\bf Keywords:} Systems out of equilibrium, random matrices, Wigner surmise.\\

\section{Introduction}
In random matrix theory the analytic expressions for the spacing distribution functions of eigenvalues $p^{(n)}(s)$ in the circular and Gaussian orthogonal ensembles (COE and GOE respectively) in the limit of large matrices are given in terms of the eigenvalues $\mu_i$ and eigenfunctions $f_i(x)$ of the following integral equation, see Ref.~\cite{mehta}:

\begin{equation}
\mu_i f_i(x)=\int^{1}_{-1}e^{i\pi x y s/2}f_i(y)dy.
\end{equation}

The spacing distributions are calculated explicitly by using

\begin{equation}
E(2r,s)=\prod^{\infty}_{i=0}(1-\lambda_{2i})\sum_{0\leq j_1<j_2<\cdots<j_r}\prod^{r}_{i=1}\left(\frac{\lambda_{j_i}}{1-\lambda_{j_i}}\right)\times\left[1-(b_{j_1}+\cdots+b_{j_r})\right],
\end{equation}

\begin{equation}
E(2r-1,s)=\prod^{\infty}_{i=0}(1-\lambda_{2i})\sum_{0\leq j_1<j_2<\cdots<j_r}\prod^{r}_{i=1}\left(\frac{\lambda_{j_i}}{1-\lambda_{j_i}}\right)\times(b_{j_1}+\cdots+b_{j_r}),
\end{equation}
where

\begin{equation}
b_j=f_{2j}(1)\int^{1}_{-1}f_{2j}(x)dx/\int^{1}_{-1}f^{2}_{2j}(x)dx,
\end{equation}
 
\begin{equation}
\lambda_{j}=s \left|\mu_j\right|^2/4,
\end{equation}
and

\begin{equation}
p^{(n)}(s)=\frac{d^2}{ds^2}\sum^{n}_{j=0}(n-j+1)E(j,s).
\end{equation}

These expressions are difficult to manage, however in Ref.~\cite{abdul}, the authors find an excellent approximation for spacing distributions $p^{(n)}(s)$ from their well-known behavior in the limits $s\rightarrow 0$ and $s\rightarrow\infty$. This approximation is easy to use and provide an excellent fit to the exact distributions. We will use this approximation many times in this paper, because of that, we summarize now its most important aspects. 

By definition, $p^{(n)}(s)$ is the probability density that an interval of length $s$ which starts at a level contains exactly $n$ levels and the next, the $n+1$ level, is in $[s,s+ds]$. In the same way, let $F^{(n)}(s)$ be the probability that an interval of length $s$ which starts at a level, contains $n$ levels. By using this definition we can write

\begin{equation}
F^{(n)}(s)=\int^{\infty}_{s}\left(p^{(n)}(s')-p^{(n-1)}(s')\right)ds'.
\end{equation}
Additionally, let $r^{(n)}(s)$ be the probability density that an interval $[0,s]$ which starts at a level at $s=0$ is limited by a level on its right side, under the condition that there are exactly $n$ levels in the interval $\left(0,s\right)$, i.e., $r^{(n)}(s)$ is the conditional probability
\begin{equation}
r^{(n)}(s)=\frac{p^{(n)}(s)}{F^{(n)}(s)},
\end{equation}
this probability is called level repulsion function. Following Ref.~\cite{abdul}, in the limit $s\rightarrow 0$, this equation can be written as

\begin{equation}\label{pitera}
p^{(n)}(s)=r^{(n)}(s)\int^{s}_{0}p^{(n-1)}(s')ds'.
\end{equation}

In the GOE ensemble the matrix elements are chosen using a Gaussian distribution, this fact suggest that $p^{(n)}(s)$ decays as Gaussian function. The appropriate function for fit is\begin{equation}\label{surmise}
p^{(n)}(s)=A_{n}s^{\alpha_{n}}e^{-B_{n}s^2},
\end{equation}
under the surmise $r^{(n)}(s)\rightarrow s^{n+1}$ with $s\rightarrow 0$. Additionally, the functions $p^{(n)}(s)$ satisfy the normalization conditions

\begin{equation}\label{pcond1}
\int^{\infty}_{0}p^{(n)}(s)ds=1,
\end{equation}
and
\begin{equation}\label{pcond2}
\int^{\infty}_{0}s p^{(n)}(s)ds=1.
\end{equation}

By using the surmise for the level repulsion and the normalization conditions, is straightforward to find \cite{abdul}

\begin{equation}\label{a}
A_{n}=2\frac{B_{n}^{(\alpha_{n}+1)/2}}{\Gamma\left(\frac{\alpha_{n}+1}{2}\right)},
\end{equation}

\begin{equation}\label{b}
B_{n}=\left[\frac{\Gamma\left(\frac{\alpha_{n}}{2}+1\right)}{(n+1)\Gamma\left(\frac{\alpha_{n}+1}{2}\right)}\right]^2,
\end{equation}

where

\begin{equation}
\alpha_{n}=n+\frac{(n+1)(n+2)}{2}.
\end{equation}

Then, the approximate spacing distribution functions $p^{(n)}(s)$ are given explicitly by
\begin{equation}\label{pwignersurmise}
p^{(n)}(s)=\left[\frac{\Gamma\left(\frac{\alpha_{n}}{2}+1\right)}{(n+1)}\right]^{\alpha_{n}+1}\frac{2 s^{\alpha_{n}}}{\Gamma\left(\frac{\alpha_{n}+1}{2}\right)^{\alpha_{n}+2}}e^{-\left[\frac{\Gamma\left(\frac{\alpha_{n}}{2}+1\right)}
{(n+1)\Gamma\left(\frac{\alpha_{n}+1}{2}\right)}\right]^2 s^2}.
\end{equation}

\begin{figure} [!htp]
\begin{center}
\includegraphics[scale=0.8]{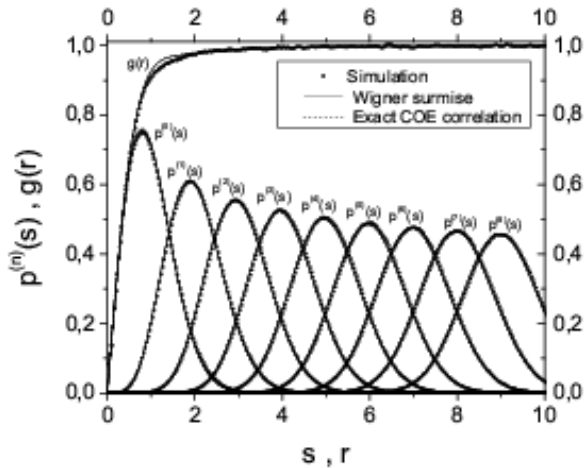}
\end{center}
\caption{Comparison between the generalized Wigner surmise and the COE ensemble. In the simulation we took $20000$ matrices of size $200\times200$.}
\label{coe}
\end{figure}

The result obtained for $\alpha_{n}$ coincides with the results
obtained by using the exact expression for the spacing distribution
functions, see Ref.~\cite{mehta}.  Notice that the approximate spacing
distributions functions are characterized by the level repulsion,
normalization condition, scaling condition for the average spacing and
Gaussian decay. This approximation is called generalized Wigner
surmise and provides a very good approximation for $p^{(n)}(s)$,
because it reproduce not only the distributions behavior in the limits
$s\rightarrow 0$ and $s\rightarrow \infty$, but also reproduce their
global behavior, as we can see in figure \ref{coe}. In particular the
function with $n=0$ is called Wigner distribution. This fit allow us
calculate also the approximate pair correlation distribution
$g(r)$. For this purpose we use

\begin{equation}\label{pkycdepares1}
g(r)=\sum^{\infty}_{n=0}p^{\left(n\right)}(r)
\,,
\end{equation}
then

\begin{equation}\label{gcoeap}
g(r)=2\sum^{\infty}_{n=0}\left[\frac{\Gamma\left(\frac{\alpha_{n}}{2}+1\right)}{(n+1)}\right]^{\alpha_{n}+1}\frac{r^{\alpha_{n}}}{\Gamma\left(\frac{\alpha_{n}+1}{2}\right)^{\alpha_{n}+2}}e^{-\left[\frac{\Gamma\left(\frac{\alpha_{n}}{2}+1\right)}{(n+1)\Gamma\left(\frac{\alpha_{n}+1}{2}\right)}\right]^2 r^2}
\,.
\end{equation}
In figure \ref{coe} we can see that this is a good approximation for $g(r)$, however, it is not as useful as the Wigner surmise for $p^{(0)}(s)$ because the exact expression for $g(r)$ is well known and easy to use, see Ref.~\cite{mehta}.
 
In Ref.~\cite{gonzalez} the authors study the statistical behavior of
several out of equilibrium domain systems which evolve with formation
of domains which grow in time. For intermediate times where the size
of the domains is much smaller than the total size $L$ of the system,
the domain size distribution exhibit a dynamic scaling. The authors
studied the statistical properties of these domains in the scaling
regime. They found that the statistical behavior of those is similar
to the one in random matrices, for example, the nearest neighbor
distribution $p^{(0)}(s)$ of several out of equilibrium domain systems
is well fitted by the Wigner surmise which also describe closely the
distribution $p^{(0)}(s)$ in the case of the circular and Gaussian
orthogonal ensembles in random matrix theory (actually this
distribution is exact in the case of $2\times2$ matrices). However,
the next distributions $\left(n>0\right)$ for these systems are
different from their counterpart in random matrix theory. Another
important aspect is the pair correlation function $g(r)$ which, in COE
and GOE ensembles and the coalescing random walk and interacting
random walk does not have any oscillation but in other systems $g(r)$
describe one oscillation near to $r=1$. For more information see
Refs.~\cite{gonzalez,ben,cornell,mettetal}. In most of the non
equilibrium domain systems, the main problem is the absence of
analytical expressions for the spacing and correlation
functions. Then, the question is: can the generalized Wigner surmise
provide a good approximation for $p^{(n)}(s)$ and $g(r)$ in the
domain systems as it happens with the random matrix ensembles?

\section{Wigner surmise for domains systems}
For all systems considered in this paper $p^{(0)}(s)$ is well described by the Wigner distribution, because of that and following the method used in the random matrix theory we propose the next model  
\begin{equation}\label{alfagen}
\alpha_n=\left\{
\begin{tabular}{cc} 
1 & for $n=0$\\
$h(n)$ & for $n\geq1$\\
\end{tabular}
\right.
\end{equation}
with $h(n)$ is a function to determine. The spacing distribution functions in this model are given by
\begin{equation}\label{pgen}
p^{(n)}(s)=\left\{
\begin{tabular}{cc} 
$\frac{\pi}{2}se^{-\frac{\pi}{4} x^2}$ & if $n=0$\\
$A_n s^{\alpha_n}e^{-B_n x^{\beta_n}}$ & if $n\geq1$\\
\end{tabular}
\right.\end{equation}
using (\ref{pcond1}) and (\ref{pcond2}), we find
\begin{equation}\label{agen}
A_n=\frac{\beta_n B_n^{\frac{1+\alpha_n}{\beta_n}}}{\Gamma(\frac{1+\alpha_n}{\beta_n})},
\end{equation}
and
\begin{equation}\label{bgen}
B_n=\left(\frac{\Gamma(\frac{2+\alpha_n}{\beta_n})}{(1+n)\Gamma(\frac{1+\alpha_n}{\beta_n})}\right)^{\beta_n}.
\end{equation}

\subsection{Independent interval approximation model (IIA)}
The independent intervals are used as an approximate solution in many equilibrium and non equilibrium systems \cite{gonzalez,salsburg,alemany} in order to find analytical results. In this approximation, $p^{\left(n\right)}(s)$ is given by the convolution product of $n+1$ nearest neighbor distribution factors, because of that, the spacing distribution functions can be calculated by using the Laplace transformation, see Ref.~\cite{salsburg}. In particular, in Ref.~\cite{gonzalez} the IIA is used to find an approximate model for the statistical behavior of two non equilibrium systems which will be explained in next sections.
\subsubsection{Independent interval model for small values of $s$}
In Ref.~\cite{gonzalez} the authors choose $p^{(0)}(s)$ equal to the
Wigner distribution. In order to apply the method of the last section,
we need to know the behavior of $p^{(n)}(s)$ for small and large
values of $s$. For the first region we expand the Wigner distribution in power series 
\begin{equation}
p^{(0)}(s)=\frac{\pi}{2}s e^{-\frac{\pi}{4}s^2}=\frac{\pi}{2}s\left(1-\frac{\pi}{4}s^{2}+\cdots\right),
\end{equation}
then, to the first order, the nearest neighbor distribution $p^{(0)}(s)$ has a lineal behavior, given by
\begin{equation}
p^{(0)}(s)\propto s.
\end{equation}

In the same limit $s\rightarrow0$, by using the independent interval
approximation for arbitrary values of $n$, we have
\begin{equation}
p^{(n)}(s)\propto \int_{0<x_1<x_2\cdots<x_n<s} x_1(x_2-x_1)\cdots(s-x_n)dx_1\cdots dx_n,
\end{equation}
which can be evaluated by using the Laplace transform 
\begin{equation}
  \widetilde{p}^{(n)}(t)\propto \frac{1}{t^{2(n+1)}},
\end{equation}
and then, taking its inverse
\begin{equation}\label{exp}
p^{(n)}(s)\propto s^{2n+1}.
\end{equation}

As consequence, in the IIA case the exponent $\alpha_n$ depends linearly on $n$ 
\begin{equation}\label{alfaida}
\alpha_{n}=2n+1.
\end{equation} 

By using this result it is possible to determine the behavior of the level repulsion function for $s\rightarrow 0$. Following Ref.~\cite{abdul} we have
\begin{equation}
r^{(n)}(s)\propto s^{f(n)},
\end{equation}
where $f(n)$ is the function to determine. By using equation (\ref{pitera}), we can write

\begin{equation}
p^{(n)}(s)\propto s^{f(n)}\int^{s}_{0}p^{(n-1)}(s')ds',
\end{equation}
then
\begin{equation}\label{fn}
p^{(n)}(s)\propto s^{f(n)+\cdots +f(0)+n}.
\end{equation}
By comparing (\ref{exp}) with (\ref{fn}) is straightforward to find
\begin{equation}
f(n)=1,
\end{equation}
for all $n\geq 0$, as a consequence
\begin{equation}\label{alfaida2}
r^{(n)}(s)\propto s,\quad s\to 0\,,
\end{equation} 
then, the level repulsion does not depend on $n$ as it happens in the COE/GOE case. 

\subsubsection{Independent interval model for large values of $s$}
Now, we need the behavior of $p^{(n)}(s)$ for large values of $s$. The exact expression for $p^{(n)}(s)$ is
\begin{equation}
p^{(n)}(s)=\int_{0<x_1<x_2\cdots<x_n<s} p^{(0)}(x_1)p^{(0)}(x_2-x_1)\cdots p^{(0)}(s-x_n) dx_1\cdots dx_n\,.
\end{equation}
In our case $p^{(0)}(s)$ is given by the Wigner surmise, then
\begin{equation}\label{IIAwig}
p^{(n)}(s)=\left(\frac{\pi}{2}\right)^{n+1}\int_{0<x_1<x_2\cdots<x_n<s} x_1 e^{-\frac{\pi}{4}x{_1}^2}\cdots(s-x_n) e^{-\frac{\pi}{4}(s-x_n)^2}dx_1\cdots dx_n.
\end{equation}
We can calculate the behavior of these functions for arbitrary values
of $n$ in this limit $s\rightarrow\infty$ as we show next. From
Ref.~\cite{gonzalez} we know that at least the first two spacing
distribution functions decay like Gaussian functions, then, we assume
that for arbitrary values of $n$ these functions have the form
$p_{asy}^{(n)}(s)=M_n s^{\gamma_n} e^{-N_n s^2} $ in the limit
$s\rightarrow\infty$. In order to eliminate the integrals in equation
(\ref{IIAwig}) we use the Laplace transformation
\begin{equation}
\label{Laplace-Wigner}
\widetilde{p}^{(n)}(l)=\left(1-l e^{l^2/\pi} \mathrm{erfc}\left(\frac{l}{\sqrt{\pi}}\right)\,\right)^{n+1},
\end{equation}
where $\erfc(z)=(2/\sqrt{\pi})\int_{z}^{\infty} e^{-t^2}\,dt$ is the
complementary Gaussian error function. In the same way we take the
Laplace transform in $p_{asy}^{(n)}$. Additionally, we expand both
transformations in Taylor series around $l=0$. Let be $Z_j$ the
$j^{th}$ coefficient in the expansion of equation
(\ref{Laplace-Wigner}) and $Y_j$ is the one for the Laplace transform
of $p_{asy}^{(n)}(s)$. We find that the coefficients of both
expansions satisfy the relation $Y_{i}/Z_{i}=Y_{j}/Z_{j}$ in the limit
$i,j\rightarrow\infty$. By using this method we can find $M_n$, $N_n$
and $\beta_n$. If fact we find that $N_n=\frac{\pi}{4 n}$ and
$\gamma_n=n+1$. In general, if we know $p^{(0)}(s)$ we can calculate
the asymptotic behavior of $p^{(n)}(s)$ under the assumption that the
IIA is valid for $s\rightarrow\infty$, but, as we will see in next
sections, this is not true always.

\begin{figure} [!htp]
\begin{center}
\includegraphics[scale=0.8]{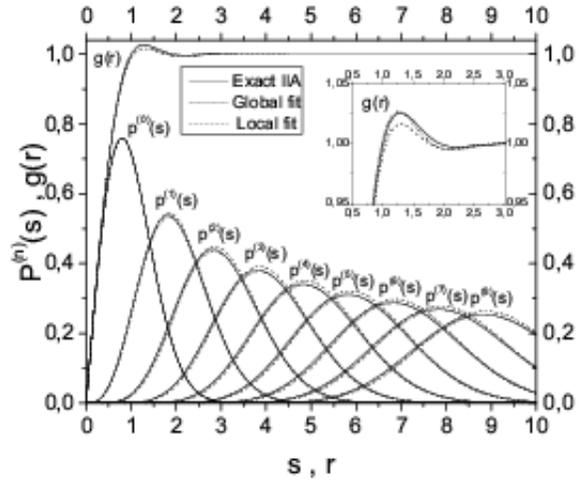}
\end{center}
\caption{Comparison between the exact statistical behavior of IIA, the generalized Wigner surmise (local fit) and the global fit.}
\label{ida}
\end{figure}

\begin{figure} [!htp]
\begin{center}
\includegraphics[scale=0.8]{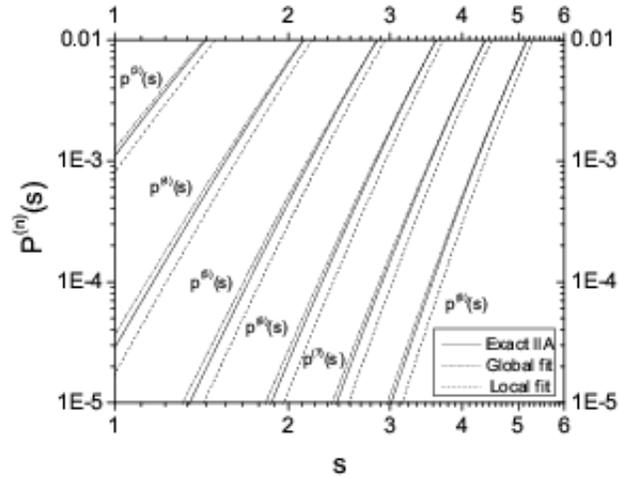}
\end{center}
\caption{Log-Log graphic for the spacing distribution functions for IIA.}
\label{ida1}
\end{figure}

\begin{figure} [!htp]
\begin{center}
\includegraphics[scale=0.8]{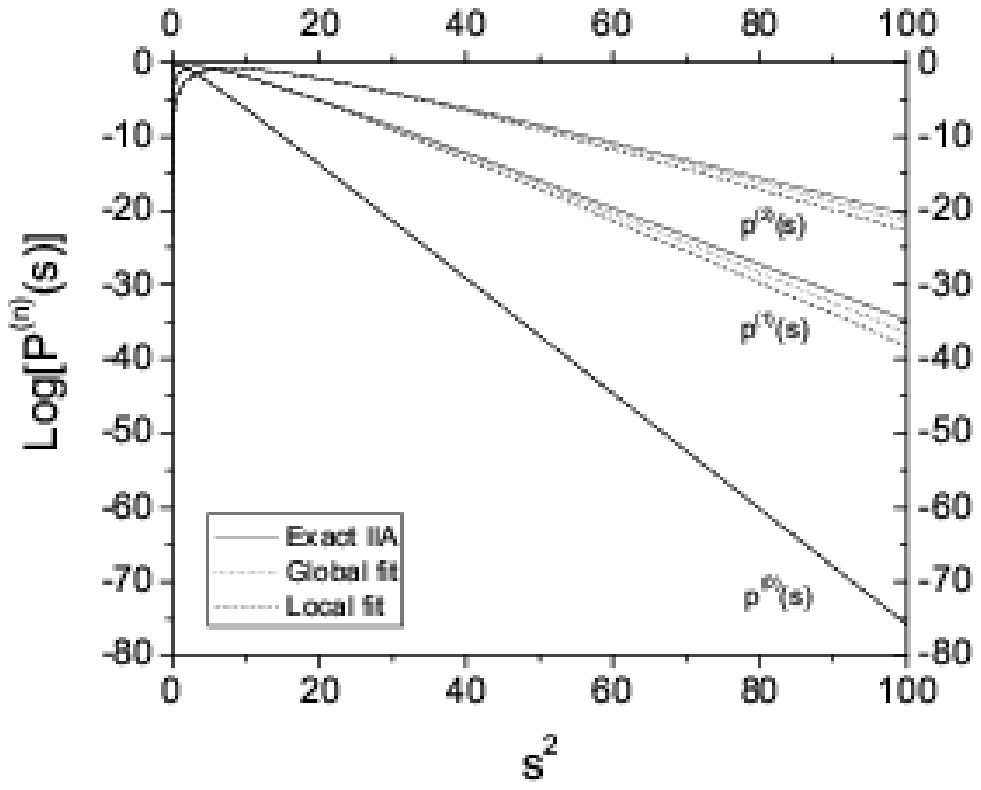}
\end{center}
\caption{Asymptotic behavior for $s\rightarrow\infty$ of the spacing distribution functions.}
\label{ida2}
\end{figure}

In the figure \ref{ida} we compare the exact statistical behavior of IIA with the generalized Wigner surmise, i.e., with a fit developed by using the behavior of $p^{(n)}(s)$ in the limits $s\rightarrow 0$ and $s\rightarrow \infty$, because of that from now on we will call it local fit. Also, we compare the global fit which was developed by using equations (\ref{alfagen}) to (\ref{bgen}) and the complete behavior of $p^{(n)}(s)$ in the interval $[0,\infty]$. By using the values of $\alpha_n$ found in the global fit, we developed a new fit to determine the global behavior of $\alpha_n$, explicitly in this case we have
\begin{equation}\label{alfaidaglobal}
\alpha_n=1.8268 n+0.9954,
\end{equation} 
this result is close to the exact exponent (\ref{alfaida}), even when we use wrong functions in the fit; for example, the exact result for $p^{(1)}(s)$ is, see Ref.~\cite{gonzalez}

\begin{equation}\label{IIAp1}
p^{\left(1\right)}(s)=\frac{\pi}{16}e^{-\frac{\pi s^2}{4}}\left(4 s +\sqrt{2}e^{\frac{\pi s^2}{8}}\left(-4+\pi s^2\right)\mathrm{erf}\left(\frac{1}{2}\sqrt{\frac{\pi}{2}}s\right)\right),
\end{equation}
which is very different form our surmise, however, both functions
(\ref{pgen}) and (\ref{IIAp1}) have the same type of behavior in the
limits $s\rightarrow 0$ and $s\rightarrow \infty$. Equation
(\ref{gcoeap}) for the correlation function it is still valid in both
cases, global and local fit, we only must use equation (\ref{alfaida}) and
(\ref{alfaidaglobal}) respectively. The main problem in the global fit
approximation it is the use of not integer exponents in the level
repulsion. Figure \ref{ida1} show the differences between the three
cases for small values of $s$, naturally in this region the graph of
the global fit is not parallel to graph of the exact result as it
actually happens in the local fit approximation. In figure \ref{ida2}
we can see the linear behavior of $p^{(n)}(s)$ in limit
$s\rightarrow\infty$, which implies that the distribution functions
decay like a Gaussian function as it was to be expected.

\subsection{Coalescing random walk (CRW)}
In the coalescing random walk the particles describe independent random walks along a one dimensional lattice and they are subjected to the reaction $A+A\rightarrow A$. This system is well studied \cite{ben,doering,ben0,ben1} and its analytical solution is well know, because of that is used as approximation to more complex systems. Let $q^{\left(n\right)}(s)$ be the conditional probability that given one particle its next neighbor is at a distance of $s$. From its definition $q^{\left(n\right)}(s)$ is given by

\begin{equation}\label{pn}
q^{\left(n\right)}(s)=\int_{0<y_1,\cdots,<y_n<s}\omega^{\left(n+2\right)}(y_1,\cdots,y_n,s)dy_1,\cdots,dy_n,
\end{equation}
with
\begin{equation}\label{omega}
\omega^{\left(n\right)}(x_1,\cdots,x_n)=-\left.\frac{\partial^n E^{\left(n-1\right)}(x_1,y_1,\cdots,x_{n-1},y_{n-1})}{\partial x_1\cdots \partial x_{n-1}\partial y_{n-1}}\right|_{y_1=x_2,\cdots,\,y_{n-1}=x_n},
\end{equation}

\begin{equation}\label{en}
E^{(n)}(x_1,y_1,\cdots,x_n,y_n,t)=\sum^{(2n-1)!!}_{p=1}\sigma_p E^{(1)}(z_{1,p},z_{2,p},t)\cdots E^{(1)}(z_{2n-1,p},z_{2n,p},t),
\end{equation}
where $z_{1,p}, z_{2,p}, . . . , z_{2n,p}$ symbolize an ordered permutation, $p$, of the variables $x_1, y_1, . . . , x_n, y_n$, such that

\begin{equation}
z_{1,p} < z_{2,p}, z_{3,p} < z_{4,p}, \cdots , z_{2n-1,p} < z_{2n,p},
\end{equation}
and

\begin{equation}
z_{1,p} < z_{3,p} < z_{5,p} \cdots < z_{2n-1,p}.
\end{equation}

The function $E^{(1)}(x_1,y_1,t)$ is the probability that from $x_1$ to $y_1$ the lattice is empty at time $t$. Then it is possible generate the complete solution for the CRW from $E^{(1)}(x_1,y_1,t)$, which is given by the solution of the diffusion equation under the suitable boundary conditions (see Ref.~\cite{ben}). In fact, the exact expression for this function is 
\begin{equation}\label{e1}
E^{(1)}(x_1,y_1,t)=\mathrm{erfc}\left(\frac{y_1-x_1}{\sqrt{8 D t}}\right),
\end{equation}
with $D$ the diffusion constant and $t$ the time, for additionally
information see Ref.~\cite{ben}. For practical purposes, the solution
given by equations (\ref{pn}) to (\ref{e1}) is hard to evaluate for
arbitrary values of $n$ but it can be evaluated in the limit $s\rightarrow 0$ using Taylor series. The case $n=0$ is trivial, the Taylor expansion for equation (\ref{e1}) is
\begin{equation}\label{expane1}
E^{(1)}(x_1,y_1,t)=1-\frac{y_1-x_1}{\sqrt{2 \pi}(D t)^{1/2}}+\frac{(y_1-x_1)^3}{24\sqrt{2 \pi}(D t)^{3/2}}-\frac{(y_1-x_1)^5}{640\sqrt{2 \pi}(D t)^{5/2}}+O(x,y)^7,
\end{equation}
then
\begin{equation}
q^{\left(0\right)}(x_2,x_1)=\omega^{\left(2\right)}(x_1,x_2)=-\left.\frac{\partial^2}{\partial x_1\partial y_1}E^{(1)}(x_1,y_1,t)\right|_{y_1=x_2},
\end{equation}
\begin{equation}
q^{\left(0\right)}(x_2,x_1)=\frac{x_2-x_1}{4\sqrt{2\pi}(D t)^{3/2}}-\frac{(x_2-x_1)^3}{32\sqrt{2\pi}(D t)^{5/2}}+O(x)^5.
\end{equation}
Making the variable change $s=\frac{x_2-x_1}{\sqrt{2\pi D t}}$ and taking into account that $p^{\left(0\right)}(s)=2\pi D t\,q^{\left(0\right)}(x_2,x_1)$, the product $D t$ disappears (dynamical scaling) in the above equation. Then, to first order, we have
\begin{equation}
p^{\left(0\right)}(s)=\frac{s \pi}{2}+O(s)^3.
\end{equation}
For small values of $s$, $p^{\left(0\right)}(s)$ has a linear behavior, i.e., $\alpha_0=1$. The case $n=1$ is more complicated, in fact we have
\begin{equation}
\omega^{\left(3\right)}(x_1,x_2,x_3)=-\left.\frac{\partial^3}{\partial x_1\partial x_2\partial y_2}E^{(2)}(x_1,y_1,x_2,y_2,t)\right|_{y_1=x_2,y_2=x_3},
\end{equation}
where 
\begin{eqnarray}
E^{(2)}\left(x_{1},y_{1},x_{2},y_{2},t\right) & = & E\left(x_{1},y_{1},t\right)E\left(x_{2},y_{2},t\right)\nonumber\\
& + & E\left(x_{1},y_{2},t\right)E\left(y_{1},x_{2},t\right) \nonumber\\
& - & E\left(x_{1},x_{2},t\right)E\left(y_{1},y_{2},t\right),\nonumber\\
\end{eqnarray}
then
\begin{equation}
\omega^{\left(3\right)}(x_1,x_2,x_3)=\frac{(x_2-x_1)(x_3-x_1)(x_3-x_2)}{32\pi (Dt)^3}+O(x)^4,
\end{equation}
in that way $q^{\left(1\right)}(x_1,x_3,t)$ is given by
\begin{equation}
q^{\left(1\right)}(x_3,x_1,t)=\int^{x_3}_{x_1}\frac{(x_2-x_1)(x_3-x_1)(x_3-x_2)}{32\pi (Dt)^3} dx_2+O(x)^5.
\end{equation}
Integrating
\begin{equation}
q^{\left(1\right)}(x_3,x_1,t)=\frac{(x_3-x_1)^4}{192\pi (Dt)^3}+O(x)^5.
\end{equation}
Using again the variable change, it is straigthfoward to find
\begin{equation}
p^{\left(1\right)}(s)=\frac{\pi^2 s^4}{24}+O(x)^5,
\end{equation}
we conclude that $\alpha_1=4$. In general for an arbitrary value of
$n$, we find that the first term in the expansion is
\begin{equation}
q^{\left(n\right)}(x_1,x_n,t)\propto \int^{x_n}_{x_1}\cdots \int^{x_3}_{x_1}\prod_{1\leq i<j\leq n} (x_j-x_i)dx_2\cdots dx_{n-1},
\end{equation}
therefore, the above equation has $(n+1)(n+2)/2$ different factors
which implies that the integrand is proportional to
$x_{i}^{(n+1)(n+2)/2}$. Making the integral and the usual variable
change, the final expression for small values of $s$ is proportional
to $s^{(n+1)(n+2)/2+n}$, explicitly, we have
\begin{equation}\label{alfacrw0}
\alpha_n=n+\frac{(n+1)(n+2)}{2}.
\end{equation} 

\begin{figure} [!htp]
\begin{center}
\includegraphics[scale=0.8]{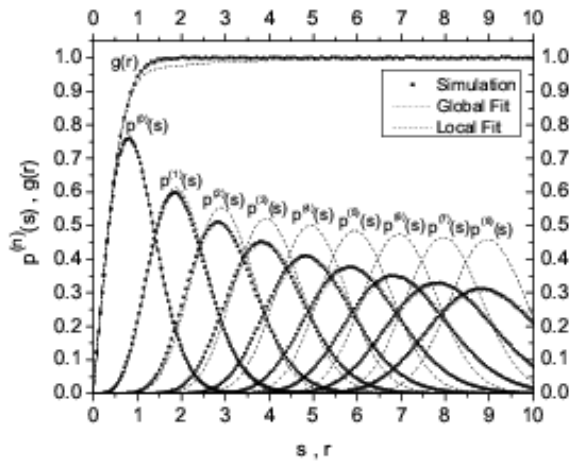}
\end{center}
\caption{Comparison between the statistical behavior of CRW, the global fit and the local fit.}
\label{crw}
\end{figure}

This is the same result reported in Ref.~\cite{abdul} for the GOE/COE case and coincides with the partial result presented in Ref.~\cite{ben1} for the CRW. We made again both fits, global and local. The global fit was made with the data from our simulation where we use a lattice with $1000$ sites and $500$ particles in $t=0$. The data to build the histograms was taken at three different times $T=50$, $T=100$ and $T=200$ over $50000$ realizations. In this case the global fit is not as accurate as in the IIA case as we can see in figure \ref{crw} but it still is a good approximation. We use again equations (\ref{alfagen}) to (\ref{bgen}); and additionally we supposed a Gaussian decay ($\beta=2$). The global fit gives

\begin{equation}\label{alfacrw}
\alpha_n=2.8688n+0.8621.
\end{equation} 

The global fit gives an erroneous exponent which depend linearly with $n$, this result it does not coincide with the analytical result (\ref{alfacrw0}), where, $\alpha_n$ is a quadratic function of $n$. The local fit it is very different from the simulation results and coincides with the statistical behavior of the COE/GOE ensembles.

\begin{figure} [!htp]
\begin{center}
\includegraphics[scale=0.8]{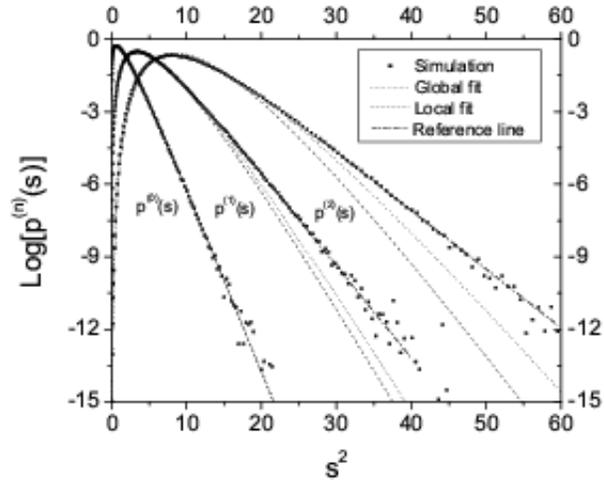}
\end{center}
\caption{Log-Log graphic for the spacing distribution functions of the CRW.}
\label{crw1}
\end{figure}

\begin{figure} [!htp]
\begin{center}
\includegraphics[scale=0.8]{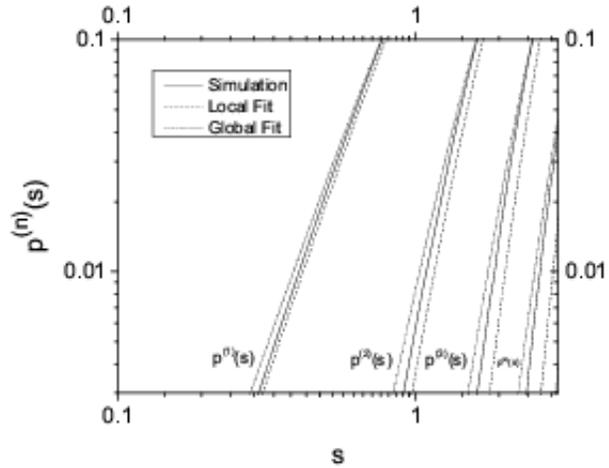}
\end{center}
\caption{Statistical behavior of CRW for small values of $s$.}
\label{crw1a}
\end{figure}

\begin{figure} [!htp]
\begin{center}
\includegraphics[scale=0.8]{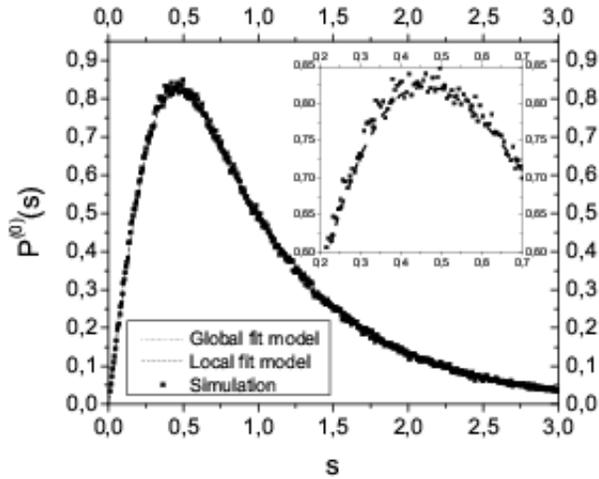}
\end{center}
\caption{Approximation for $p^{(0)}_{ann}(s)$ by using global and local fits.}
\label{crw2}
\end{figure}

\begin{figure} [!htp]
\begin{center}
\includegraphics[scale=0.8]{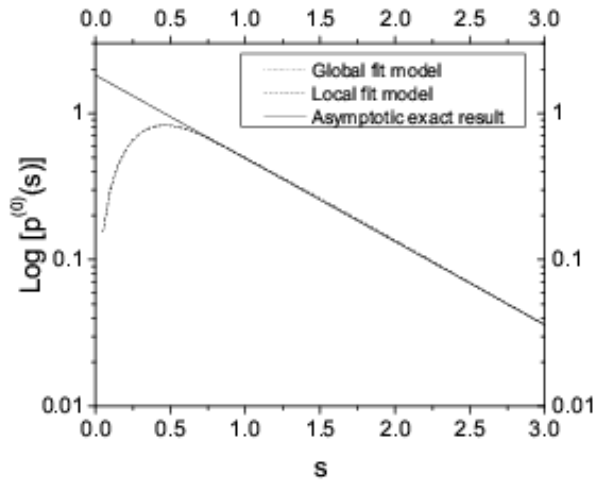}
\end{center}
\caption{Asymptotic behavior of $p^{(0)}_{ann}(s)$.}
\label{crw3}
\end{figure}

Although the global and local fit models are approximate, we can use them as a good approximations in some cases. For example, in Ref.~\cite{ben1} the authors find an exact relation for the nearest neighbor distribution $p^{(0)}_{ann}(s)$ in the annihilation random walk in terms of $p^{(n)}(s)$ of the coalescing random walk. Explicitly, they found 

\begin{equation}\label{modeloaniquilacion}
p^{(0)}_{ann}(s)=\sum_{n\geq 0}\frac{1}{2^n}p^{(n)}(2s),
\end{equation}
In order to test the validity of our approximations, we implement a simulation for the annihilation random walk for a one dimensional lattice with $2000$ sites, $100$ particles at $t=0$ over $20000$ realizations, the histogram was build by using three times $T=1000$, $T=1500$ and $T=2000$. By using the global and the local fit for the distribution functions $p^{(n)}(s)$ of the CRW, with equation (\ref{modeloaniquilacion}), we find two analytical models for the annihilation random walk. We can see in figure \ref{crw2} that the global and local fits provides a good approximation for the nearest neighbor distribution of the annihilation random walk. Additionally, figure \ref{crw3} compare global and local fit with the asymptotic result $p^{(0)}_{ann}(s)\approx1.8167 e^{-1.3062s}$ given in Ref.~\cite{ben1}.

\subsection{Spin System}
\begin{figure} [!htp]
\begin{center}
\includegraphics[scale=0.8]{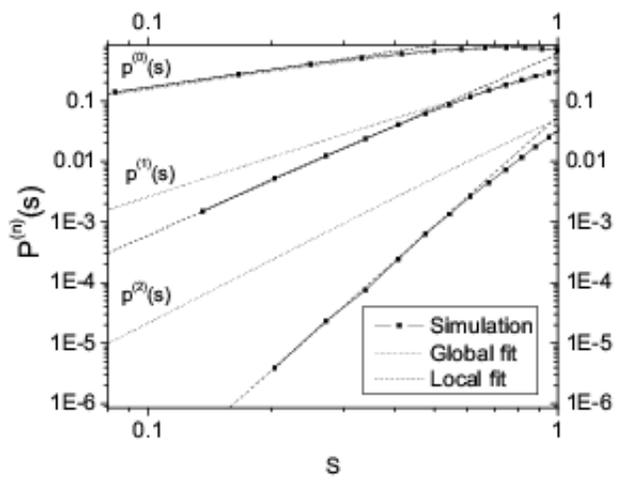}
\end{center}
\caption{Log-Log graphic for the spacing distribution functions of spin system.}
\label{espines1}
\end{figure}

This system was introduced in Ref.~\cite{cornell}, where the authors consider a chain of $L$ Ising spins with nearest neighbor ferromagnetic interaction $J$. The chain
is subject to spin-exchange dynamics with a driving force $E$ that
favors motion of up spins to the right over motion to the left. In
this case we do not have an analytical solution for the spacing
distribution functions, because of that, we must start exploring
numerically the behavior of $p^{(n)}(s)$ for small and large values of
$s$. In figure \ref{espines1}, we can see the linear behavior of the spacing distribution function for $s\rightarrow0$. Using values in this region we develop a fit which suggest that $\alpha_1=3$ and $\alpha_2=6$ approximately. Naturally $\alpha_0=1$, however it is very difficult to know using this method the next exponents because it is not possible develop a numerical simulation with enough precision. 

\begin{figure} [!htp]
\begin{center}
\includegraphics[scale=0.8]{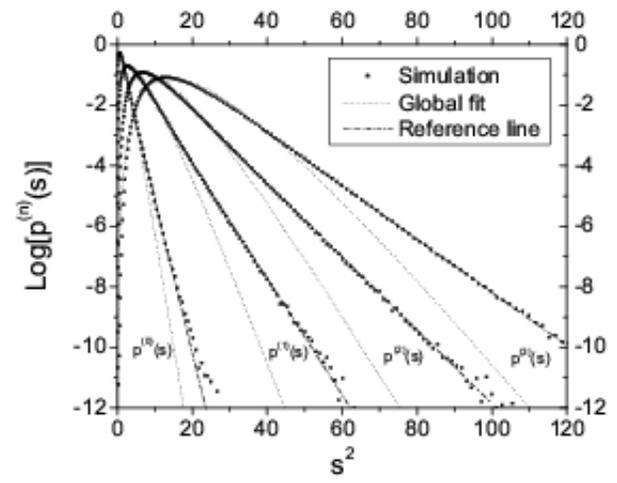}
\end{center}
\caption{Asymptotic behavior of the spin system.}
\label{espines2}
\end{figure}
Curiously, these exponents for $n=0,1,2$ are given by the equation
\begin{equation}
\alpha_n=\frac{(n+1)(n+2)}{2},
\end{equation}
which is very similar to its counterpart in COE and GOE cases. For $s\rightarrow\infty$, $p^{(n)}(s)$ decay like a Gaussian function as we can see in figure \ref{espines2}. In this case the global fit gives

\begin{equation}\label{alfaespines}
\alpha_n=1.270 n + 0.920.
\end{equation} 

\begin{figure} [!h]
\begin{center}
\includegraphics[scale=0.8]{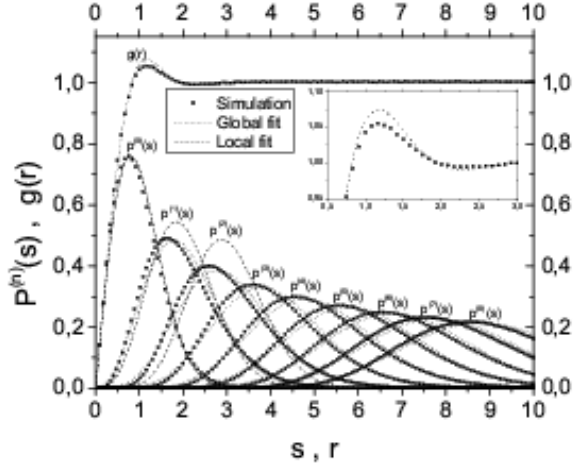}
\end{center}
\caption{Comparison between the statistical behavior of the spin system and the global fit.}
\label{espines}
\end{figure}

In figure \ref{espines}, we show the results given by equations (\ref{alfagen}) to (\ref{bgen}) for the global fit in comparison with the simulation results which was made with a lattice with $1000$ sites, equal number of spins up and down taken at two times $t=34$ and $t=48$ to build the histograms. The result for $g(r)$ is very good with a maximum error of $2.5\%$. Unfortunately this approximation is not good enough for $p^{(n)}(s)$ but at least it reproduce qualitatively the behavior of the real functions for $s\rightarrow\infty$. The local fit gives terrible results as it happens in the CRW case.

\subsection{Gas System}
\begin{figure} [!htp]
\begin{center}
\includegraphics[scale=0.8]{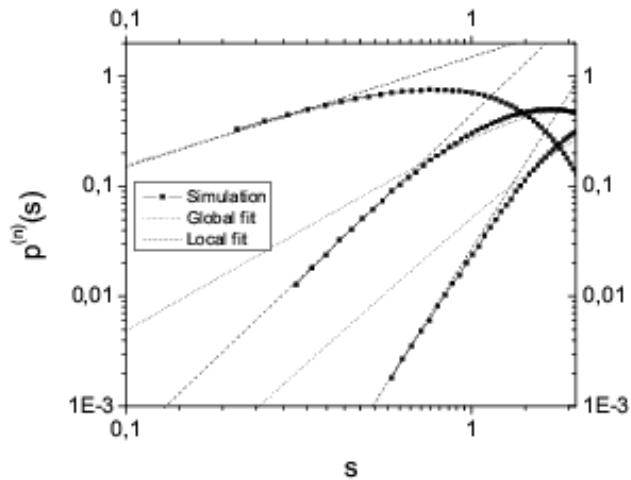}
\end{center}
\caption{Log-Log graphic for the spacing distribution functions of gas system.}
\label{gas}
\end{figure}

\begin{figure}[htp]
\begin{center}
$\begin{array}{cc}
\includegraphics[scale=0.5]{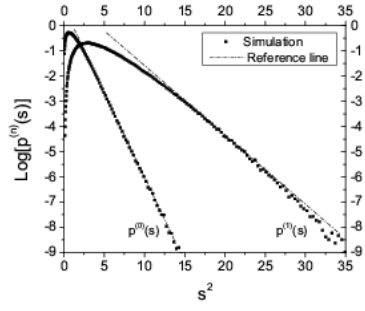} & \includegraphics[scale=0.5]{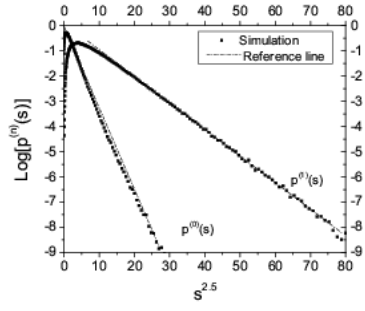}\\
(a) & (b) \\
\includegraphics[scale=0.5]{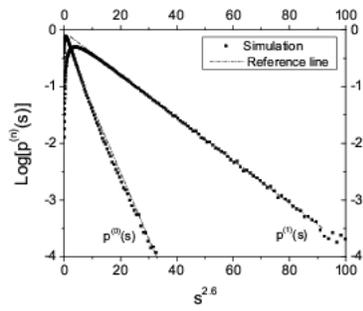} & \includegraphics[scale=0.5]{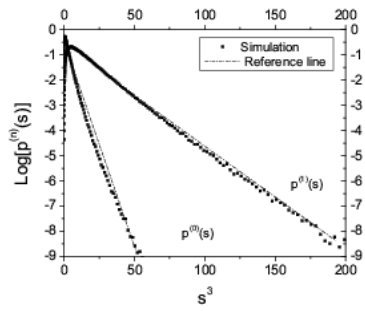}\\
(c) & (d) \\
\end{array}$
\end{center}
\caption{Asymptotic behavior of $p^{(0)}(s)$ and $p^{(1)}(s)$ for the gas system.}
\label{gastres}
\end{figure}

\begin{figure} [!htp]
\begin{center}
\includegraphics[scale=0.8]{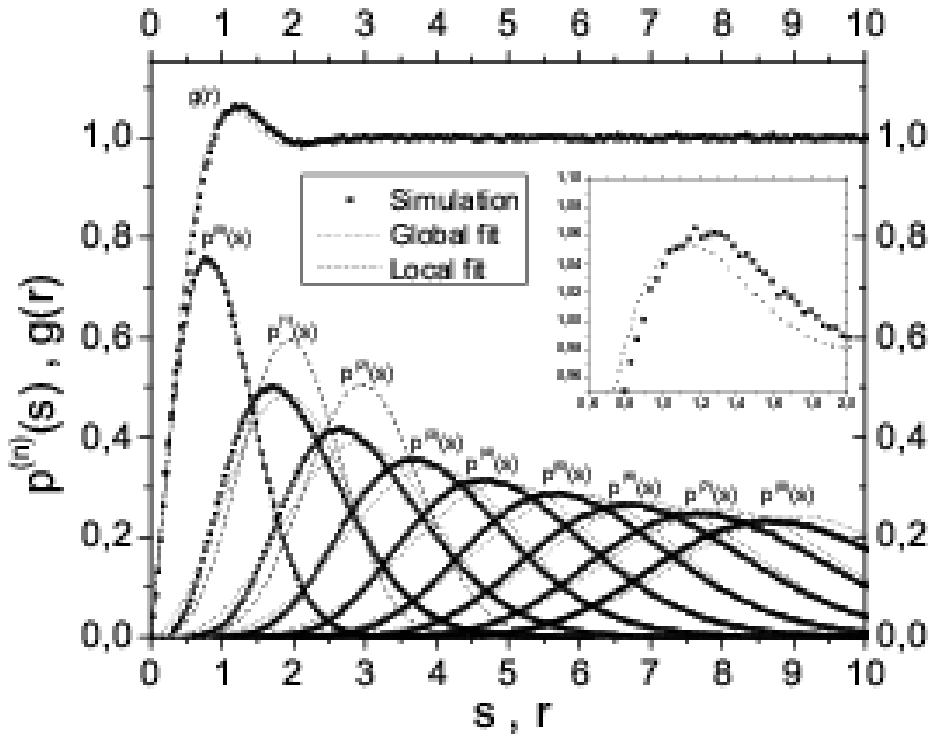}
\end{center}
\caption{Comparison between the statistical behavior of the gas system and the global fit.}
\label{gas2}
\end{figure}

This system was originally studied in \cite{mettetal}. There, the
authors studied the biased diffusion of two species in a fully
periodic $2\times L$ rectangular lattice half filled with two equal
number of two types of particles (labeled by their charge $+$ or
$-$). An infinite external field drives the two species in opposite
directions along the $x$ axis (long axis). The only interaction
between particles is an excluded volume constraint, i.e., each lattice
site can be occupied at most by only one particle. As it happens in
the spin system, we do not know an analytic solution for the spacing
and pair correlation functions. We follow the same method used in the
spin system. In figure \ref{gas}, we can see the linear behavior of
$p^{(n)}(s)$ which, by fit, give us $\alpha_1=3$ and $\alpha_2=5$
approximately, and of course $\alpha_0=1$. This fact suggest a linear
behavior for $\alpha_n$ given by
\begin{equation}
\alpha_n=2n+1 
\end{equation}
but again we could not find the next exponents with enough precision
in order to validate the above equation. For $s\rightarrow\infty$ we
found that $p^{(0)}(s)$ decays like a Gaussian function $(\beta=2)$,
but for $n>0$, we found that $\beta$ is an indeterminate function of
$n$. For example in figure \ref{gastres} we can see the asymptotic
behavior for two consecutive spacing distribution functions, the
figure suggest $\beta=2$ for $p^{(0)}(s)$ and $\beta\neq2$ for
$p^{(1)}(s)$ as it happens in Ref.~\cite{aarao}. Because it is
difficult determine the exact value of $\beta$ from the graphics, we
implement a linear regression to find which value of $\beta$ give us a
better "straight" line. With this method we find for example, that
$\beta=2.6$ for $n=1$, $\beta=3$ for $n=5$ and $\beta=3.2$ for
$n=8$. In the linear regressions we took values between $5.5\geq s
\geq2.5$, $11.7\geq s \geq7$ and $15.5\geq s \geq10$
respectively. Because of that, for the gas system we propose a model
where $\beta$ depends on $n$. In particular we choose
$\beta=2.6+0.1(n-1)$. With this model, the global fit gives
\begin{equation}
\alpha_n=1.016 n + 0.788.
\end{equation}
The results of the global fit are show in figure \ref{gas2}, again we find good fit for $g(r)$ with a maximum error of $2\%$ approximately but the agreement for $p^{(n)}(s)$ is not so good. Additionally, we include the first spacing distribution obtained with the local fit and our model for $\beta_n$.

\section{Conclusion}
In COE and GOE ensembles, the spacing distribution functions
$p^{(n)}(s)$ can be well described by using their behavior for small and
large values of $s$ (local fit) as it happens in IIA case, however, this
is not true for more complex systems like CRW, spin and gas
systems. This result was to be expected because in general the spacing
distribution functions are characterized also by their inter medium
behavior. In general, the global fit gives better results in
comparison with the local fit but it fails to reproduce the level
repulsion, in fact, gives non integer exponents. The level repulsion
for the CRW has the same behavior that the circular and Gaussian
orthogonal ensembles, i.e., both systems are equivalents for
$s\rightarrow 0$. The numerical results suggest that the IIA and the
gas system are also equivalents in that region. We find numerical
evidence that the spacing distributions functions for gas system is
described by a non universal function, in fact, they decay as $M_n
s^{\gamma_n} e^{-N_n~s^{\beta_n}}$ for $n>0$, with $\beta_n$ an
indeterminate function of $n$. In general the global and local fit
provides a first approximation for $p^{(n)}(s)$ and $g(r)$, which can
be used as a good approximation as it happens in the annihilation
random walk case. These approximations also serve to classify the
spacing distribution functions according to their level of repulsion
and their decay functional form.

\section*{Acknowledgments}
This work was partially supported by an ECOS Nord/COLCIENCIAS action of French 
and Colombian cooperation and by the Faculty of Sciences of Los Andes University.


\end{document}